**Submitted to 'Chinese Physics C'**

# Hard X-ray optics simulation using the coherent mode decomposition of Gaussian Schell model*

HUA Wen-Qiang(滑文强)  BIAN Feng-Gang(边风刚)  SONG Li(宋丽)

LI Xiu-Hong(李秀宏)  WANG Jie(王劼)

Shanghai Synchrotron Radiation Facility, Shanghai Institute of Applied Physics, Chinese Academy of Sciences, Shanghai 201204, China

**Abstract:** The propagation of hard X-ray beam from partially coherent synchrotron source is simulated by using the novel method based on the coherent mode decomposition of Gaussian Schell model and wave-front propagation. We investigate how the coherency properties and intensity distributions of the beam are changed by propagation through optical elements. Here, we simulate and analyze the propagation of the partially coherent radiation transmitted through an ideal slit. We present the first simulations for focusing partially coherent synchrotron hard X-ray beams using this novel method. And when compared with the traditional method which assumes the source is a totally coherent point source or completely incoherent, this method is proved to be more reasonable and can also demonstrate the coherence properties of the focusing beam. We also simulate the Young's double slit experiment and the simulated results validate the academic analysis.

---

* Supported by National Natural Science Foundation of China (11105215)

1)E-mail: wangjie@sinap.ac.cn

## 1. Introduction

Coherence-based techniques, such as coherent X-ray diffraction and X-ray holography, rely on a high degree of correlation between wavefronts at different points in space and time. With poor coherence, the key image features may be blurred out. When using synchrotron X-rays, high temporal coherence is achievable with undulators and monochromators and a sufficiently high degree of spatial coherence is also important [1]. With the development of the third generation synchrotron sources and the construction of X-ray free-electron laser, the available coherent output has been growing rapidly [2].

However, the high spatial coherence, in turn, has given rise to new problems. Firstly, understanding the coherence properties of X-ray beams is of vital importance [3-5]. Correspondingly, more efforts have been made in measuring the coherence properties of the synchrotron radiation [6-20] and modeling the X-rays beams [4-6, 21-25]. Secondly, various optical components in the beamline, such as slits, beryllium windows and mirrors, become the origins of interference fringes when their compositional or structural homogeneity or both are insufficient [26]. It also raises questions concerning the effect of such optics on the coherence of the beam and whether the coherence can be usefully transported to the experimental apparatus [26-33]. From the preliminary analysis, it is clear that an effective and useful tool for describing and calculating the coherence properties of X-ray sources and their beams which pass through different optical elements is in high demand.

Nowadays, many methods have been developed to calculate the beam profile at the sample position. And most calculations are based on the ray tracing or wave propagation approach, but both are limited when the radiation is neither fully coherent nor totally incoherent but rather partially coherent. Recently, the Gaussian Schell-model (GSM) has been used to describe the coherence properties of undulator sources and XFEL sources [4-6, 19, 21, 24, 29]. And the simulation can accurately describe the beam properties, but the amount of calculation is very great. While the latest proposed Coherent Mode Decomposition (CMD) for GSM is more convenient and efficient [5, 6, 19, 24]. In this paper, we extend this new approach to calculate the coherence properties of X-ray beams passing through an ideal slit, focusing with an ideal thin lens and model the Young double slit experiment.

## 2. Theory

### 2.1. Fundamental concepts

In the theory of coherence, a partially coherent field is described using the second order correlations of wave field via [34, 35] the mutual coherence function (MCF) , $\Gamma(r_1, r_2; \tau)$ . It defines the correlations between the two complex values of the electric field $E(r_1, t+\tau)$ and $E(r_2, t)$ at different points $r_1$ and $r_2$ and at different time

$$\Gamma(r_1, r_2; \tau) = \langle E(r_1, t+\tau) E^*(r_2, t) \rangle_T , \tag{1}$$

where τ is the time delay and the bracket $\langle ... \rangle_T$ represents an averaging over time T much longer than the fluctuation time of the X-ray field, which also assumes that the field is ergodic and stationary.

We also need to introduce the cross spectral density (CSD) function. It is the

Fourier transform pair with the MCF in the time-frequency domain

$$W(r_1, r_2; \omega) = \int \Gamma(r_1, r_2; \tau) e^{i\omega\tau} d\tau . \quad (2)$$

Another important concept is the spectral degree of coherence, which is the normalized function of $W(r_1, r_2; \omega)$ at frequency $\omega$.

$$\mu_{12}(\omega) = \frac{W(r_1, r_2; \omega)}{\sqrt{I(r_1; \omega) I(r_2; \omega)}} \ , \ I(r; \omega) = W(r, r; \omega), \quad 0 \le |\mu_{12}(\omega)| \le 1 . \quad (3)$$

2.2. The coherent mode decomposition of Gaussian Schell model

It is often to assume that a real synchrotron source can be represented by its equivalent model that produces X-ray beam with similar statistical properties. A GSM beam is a particular type of partially coherent wave field which is usually used to describe the radiation coherence properties as well as intensity distributions. In this model the CSD in the source plane, $W_S(r_1, r_2, z = 0; \omega)$, is described by:

$$W_S(r_1, r_2, z = 0; \omega) = \left[ I(r_1) I(r_2) \right]^{1/2} \mu_s(r_2 - r_1; \omega) , \quad (4)$$

where

$$I(r) = I_{Sx} I_{Sy} \exp\left( -\frac{x^2}{2\sigma_{Sx}^2} - \frac{y^2}{2\sigma_{Sy}^2} \right) , \quad (5)$$

$$\mu_s(r_2 - r_1; \omega) = \exp\left( -\frac{(x_2 - x_1)^2}{2\xi_{Sx}^2} - \frac{(y_2 - y_1)^2}{2\xi_{Sy}^2} \right) . \quad (6)$$

Here the subscript S is used on the variables that are in the source plane. $I_{Sx,y}$ are the positive constant representing the maximum intensity in the respective directions that are set to one in this paper. The parameters $\sigma_{Sx,y}$ are root-mean-squared source size in the x and y directions and $\xi_{Sx,y}$ give the coherence length of the source.

Then, the propagation of the CSD from the source plane to the plane at different distance z is given by [35]:

$$W(u_1,u_2,z;\omega)=\int_\Sigma\int_\Sigma W_S(r_1,r_2,z=0;\omega)P_z^*(u_1-r_1;\omega)P_z(u_2-r_2;\omega)dr_1dr_2 \,. \quad (7)$$

Here, we adopt the paraxial approximation which is well satisfied by a synchrotron beamline [28]. And $P_z(u-r;\omega)$ is the propagator that describes the propagation of radiation in free space. It is defined as

$$P_z(u-r;\omega)=\frac{1}{i\lambda z}\exp\left(\frac{ik}{2z}(u-r)^2\right), \quad (8)$$

where $k=\frac{2\pi}{\lambda}$ is the wave vector and $\lambda$ is the wavelength of radiation.

According to the theory developed in Ref. [35], the CSD of partially coherent, statistically stationary field of any state of coherence can be decomposed into a sum of independent coherent modes under very general conditions [19]

$$W(r_1,r_2;\omega)=\sum_n \beta_n(\omega)E_n^*(r_1;\omega)E_n(r_2;\omega)\,, \quad (9)$$

where $\beta_n(\omega)$ are the eigenvalues and $E_n(r;\omega)$ are the eigenfunctions of the integral equation

$$\int W(r_1,r_2;\omega)E_n(r_1;\omega)dr_1=\beta_n(\omega)E_n(r_2;\omega)\,. \quad (10)$$

$E_n(r;\omega)$ are known as the coherent modes and are mutually incoherent, $\beta_n(\omega)$ describe the occupancy in each mode.

According to Eq. (9), the modes $E_n$ and their corresponding eigenvalues $\beta_n$ can be found for x and y directions respectively. The CSD in x direction can be described as:

$$W(x_1,x_2;\omega)=\sum_n \beta_n^x(\omega)E_n^*(x_1;\omega)E_n(x_2;\omega)\,. \quad (11)$$

And $E_n,\beta_n^x$ are described by the Gaussian Hermite modes [36]:

$$\frac{\beta_n^x}{\beta_0^x} = \left(\frac{2\sigma_{Sx} - \delta_{Sx}}{2\sigma_{Sx} + \delta_{Sx}}\right)^n = \left[\frac{q_x^2}{2} + 1 + q_x\sqrt{\left(\frac{q_x}{2}\right)^2 + 1}\right]^{-n} , \quad (12)$$

where, $\frac{1}{\delta_{Sx,y}^2} = \frac{1}{(2\sigma_{Sx,y})^2} + \frac{1}{\xi_{Sx,y}^2}$, $q_x = \frac{\xi_{Sx}}{\sigma_{Sx}}$ , $\beta_0^x = \frac{\sqrt{8\pi} I_{0x}\sigma_{Sx}\delta_{Sx}}{2\sigma_{Sx} + \delta_{Sx}}$,

$$E_n(x) = \left(\frac{k}{\pi z_x^{eff}}\right)^{1/4} \frac{1}{(2^n n!)^{1/2}} H_n\left(\sqrt{\frac{k}{z_x^{eff}}}x\right) \exp\left(-\frac{k}{2z_x^{eff}}x^2\right) . \quad (13)$$

Moreover, $H_n(x)$ are the Hermite polynomials of order n, $z_x^{eff} = k\sigma_{Sx}\delta_{Sx}$.

Then, the propagation of the field from the source through free space to the first optical element at the position $z_1$ for each mode can be performed by utilizing the Huygens-Fresnel principle

$$E_n(u, z_1; \omega) = \int_\Sigma E_n(r, z_0; \omega) P_z(u - r; \omega) dr . \quad (14)$$

When the hard X-ray beam passes through the optical element which can be a pair of slits, beryllium window, a lens or a mirror, the transmitted modes are given by

$$E_{out}(r, z) = T(r) E_{in}(r, z) , \quad (15)$$

where $T(r)$ characterizes the optical element's complex valued amplitude transmittance function.

Durinng the next step, the transmitted modes are propagatinging to the next optical element using Eq. (14). Finally, after propagation from the synchrotron source through all optical elements present in the beamline, each mode is calculated in the plane of observation. Then, the CSD representing the beam properties in the plane of observation is determined by Eq. (9, 11).

## 3. Application of CMD to hard X-ray synchrotron radiation optics simulation

For convenience, we use the same 5 m long undulator source PETRA III at DESY and a photon energy of 12 keV as calculated in Ref. [5]. The source parameters are shown in Table 1. Owning to the symmetry of the GSM and the high coherence in the vertical direction, we analyse the coherence properties of the hard X-ray radiation only in the vertical direction. As discussed in Ref. [25], modes with a contribution larger than 0.001 can’t be neglected, and only in this way, can the difference of GSM and CMD be too smaller to be perceived. As a result, 8 modes used in the calculations are presented here.

Table 1. Parameters of the high brilliance synchrotron radiation source PETRA III for a 5 m undulator [5] (energy E=12 keV).

| | Vertical |
|---|---|
| Source size, μm | 5.5 |
| Source divergence, μrad | 3.8 |
| Transverse coherence length at the source, μm | 4.53 |

3.1 Propagation of partially coherent beam in free space

In this section, we simulate the propagation of partially coherent beam generated by a GSM in free space. When only a small number of modes are applied to describe the CSD of the beam, can the CMD be much more convenient in the analysis of the propagation of partially coherent radiation [24]. The coherent modes separately propagate along the optical axis and the CSD can be calculated at any position. The propagation of the TEM5 mode in free space is shown in Fig. 1(a).

The beam properties are analyzed at 30m downstream the source. These are illustrated in detail in Fig. 1(b, c). The beam intensity of GSM and its CMD are shown in Fig. 1(b). And Fig. 1(c) shows modulus of the complex degree of coherence $|\mu(\Delta x)|$ of two models as a function of the separation $\Delta x$. From these figures we clearly see that the intensity difference is small, but the modulus of the complex degree of coherence mismatches at large $\Delta x$. When at large separation, the contribution of high order modes which was neglected is great, but it doesn't matter because the value of $|\mu(\Delta x)|$ at large $\Delta x$ is usually not taken into account.

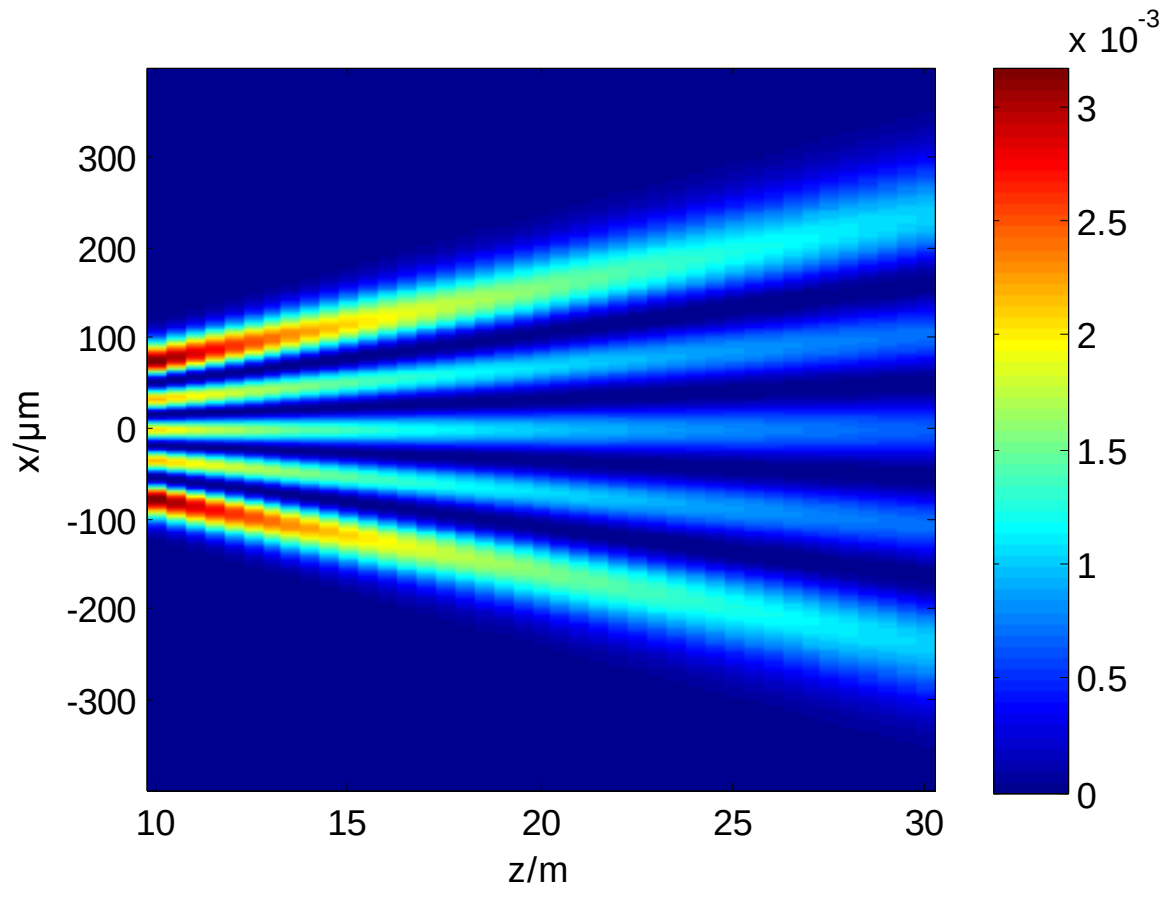

(a)

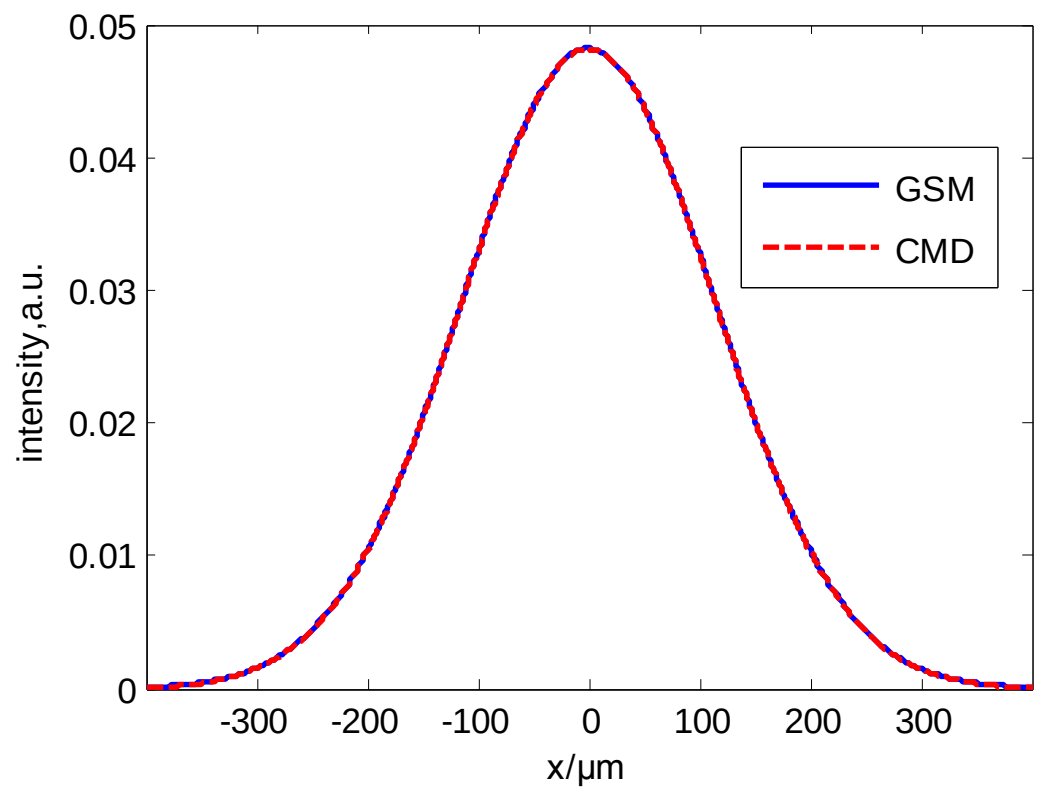

(b)

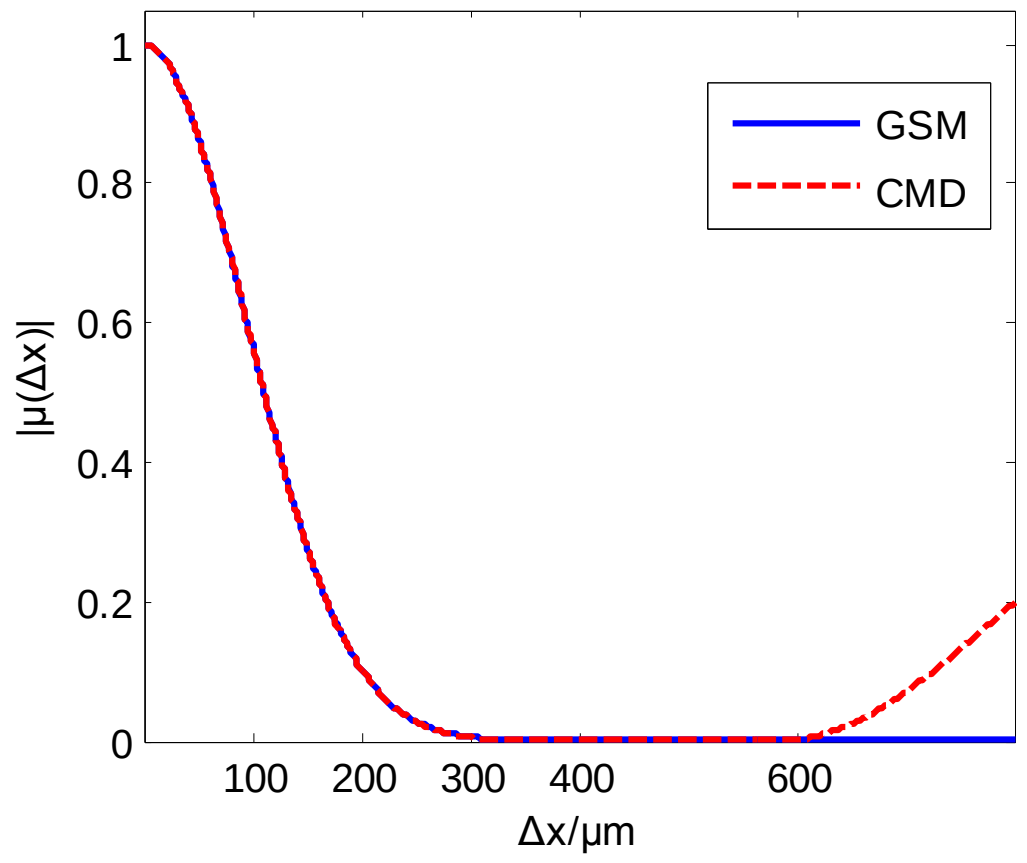

(c)

Fig. 1. (a) Propagation of the TEM5 mode in free space. (b) The intensity distribution of GSM (blue solid line) and its CMD (red dashed line). (c) The modulus of the complex degree of coherence of GSM (blue solid line) and its CMD (red dashed line) as a function of the separation $\Delta x$ .

### 3.2 Propagation of partially coherent beam through a slit

As in Ref. [37], pinholes or slits are often used for limiting the beam size or divergence when they can also pick out the coherent part of the beam for coherent experiments. In this paper, we calculate the propagation of unfocused partially coherent beam through a slit. The slit is positioned 30 m downstream the source and the beam properties are analyzed 50m downstream the slit. The single slit diffraction intensity distributions of the eight lowest modes and only the fundamental mode with different slit sizes are shown in Fig. 2(a, b, c, d). And we clearly see the transition from the Fresnel to Fraunhofer diffraction.

It is easy to know that if we want the beam more coherent, we should filter out all the high order modes and let the fundamental mode transmit alone. Since the distribution of the fundamental mode is closer to the optical axis, we can gradually reduce the slit size to filter out other modes and increase the ratio of the fundamental mode to all modes as discussed in Ref. [25]. We clearly see the change of the intensity distributions of the beam and only the coherent fundamental mode from Fig. 2(a, b, c, d). Fig. 2(e) shows the modulus of the spectral degree of coherence $|\mu(\Delta x)|$ as a function of the separation $\Delta x$ in the observation plane. And the oscillation curve in Fig. 2(e) may be caused by the oscillation of the intensity curve due to the Fresnel diffraction. It can be seen in Fig. 2(a, b, c, d) that when the slit size gradually decreases from 400 μm to 5 μm, the transmitted beam size decreases first and then increases after arriving at a minimum value [27]. Correspondingly, the values of $|\mu(\Delta x)|$ in the transmitted central beam which will be illuminated on the sample are enhanced significantly for smaller slit as is shown in Fig. 2(e), and the correlations of the transmitted beam gradually get higher. As a result of our simulations, we can decrease the slit size to get the beam with higher degree of coherence, but at the cost of the photon flux. In Fig. 2(e), we also plot the photon flux transmittance as a function of slit size. This analysis can give a hand to the beam users to count the cost as they change the slit size for more coherent beams. And it can also help the beamline designer to optimize the slits' position and size for coherent-based experiments.

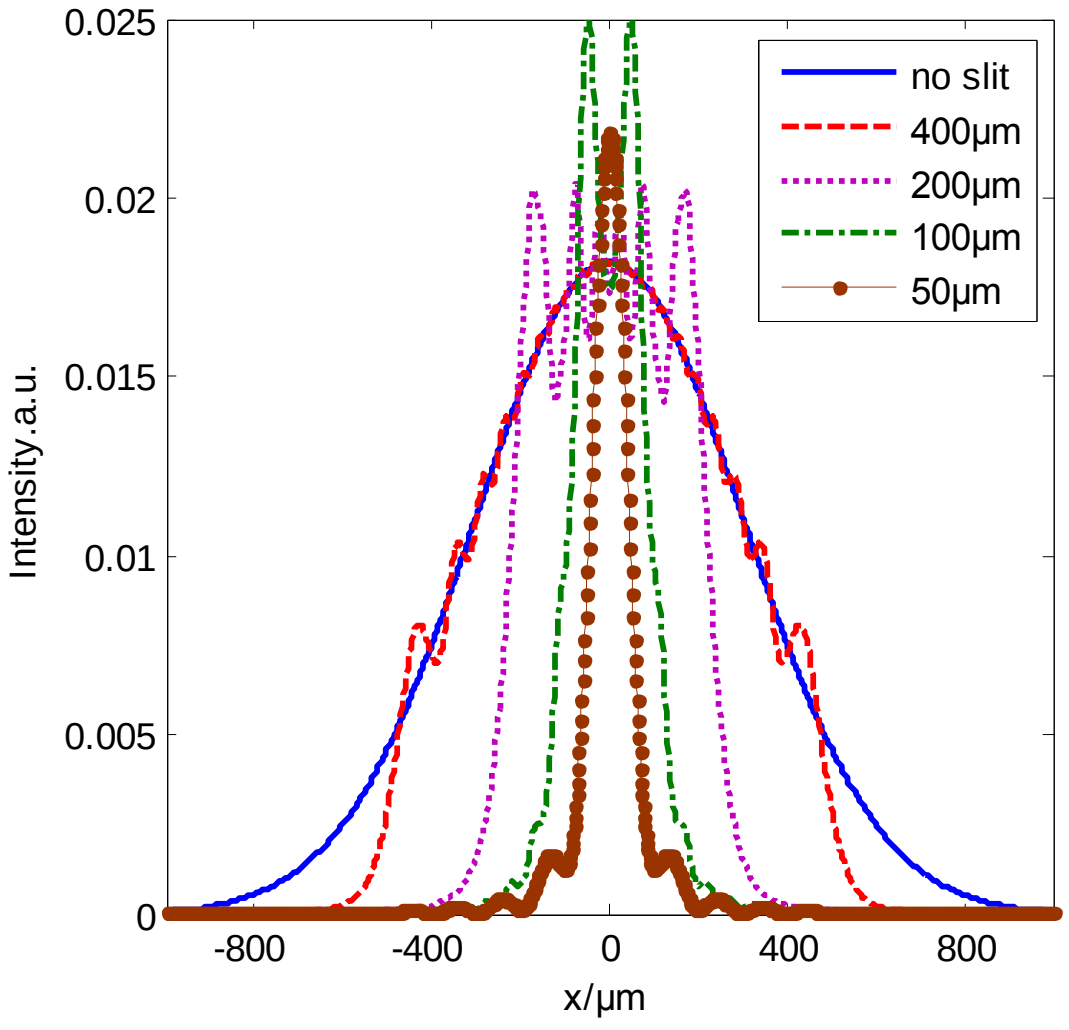

**(a)**

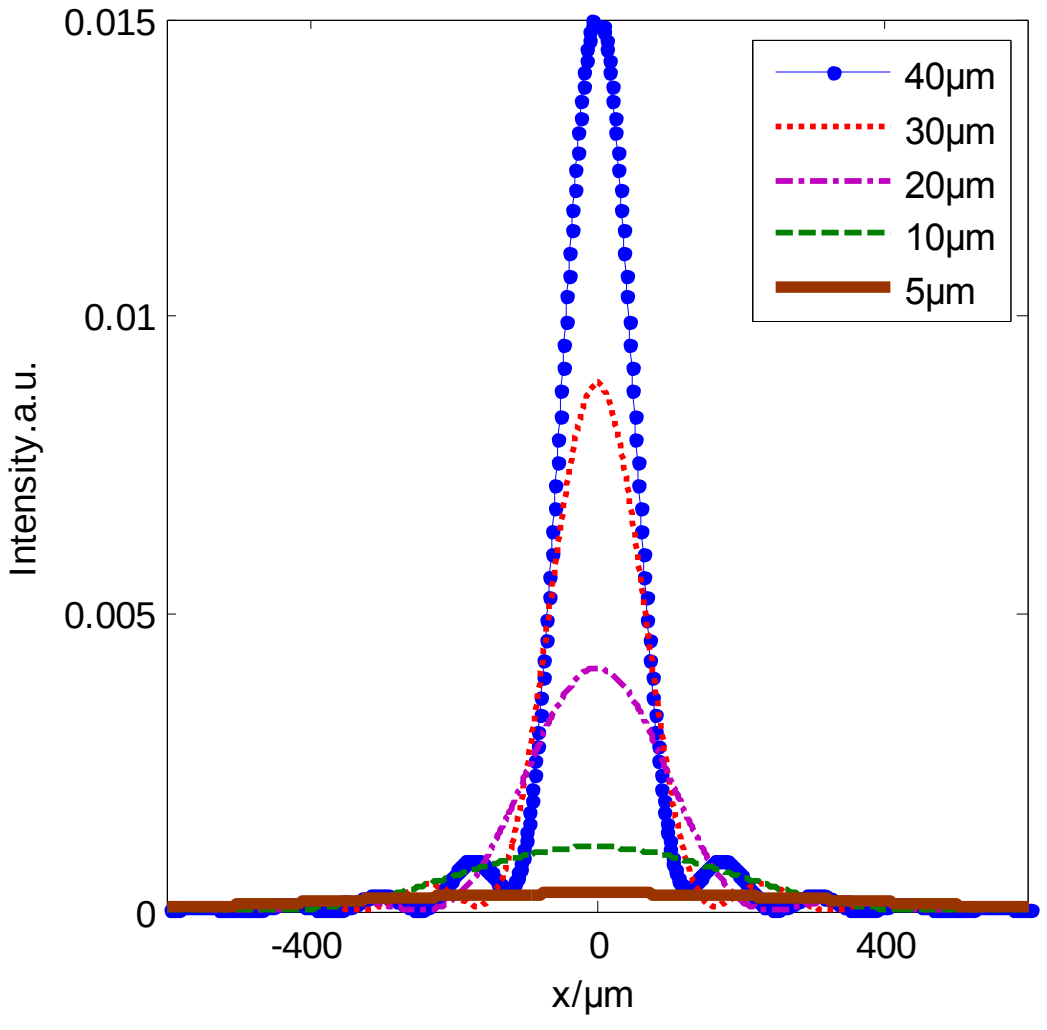

**(b)**

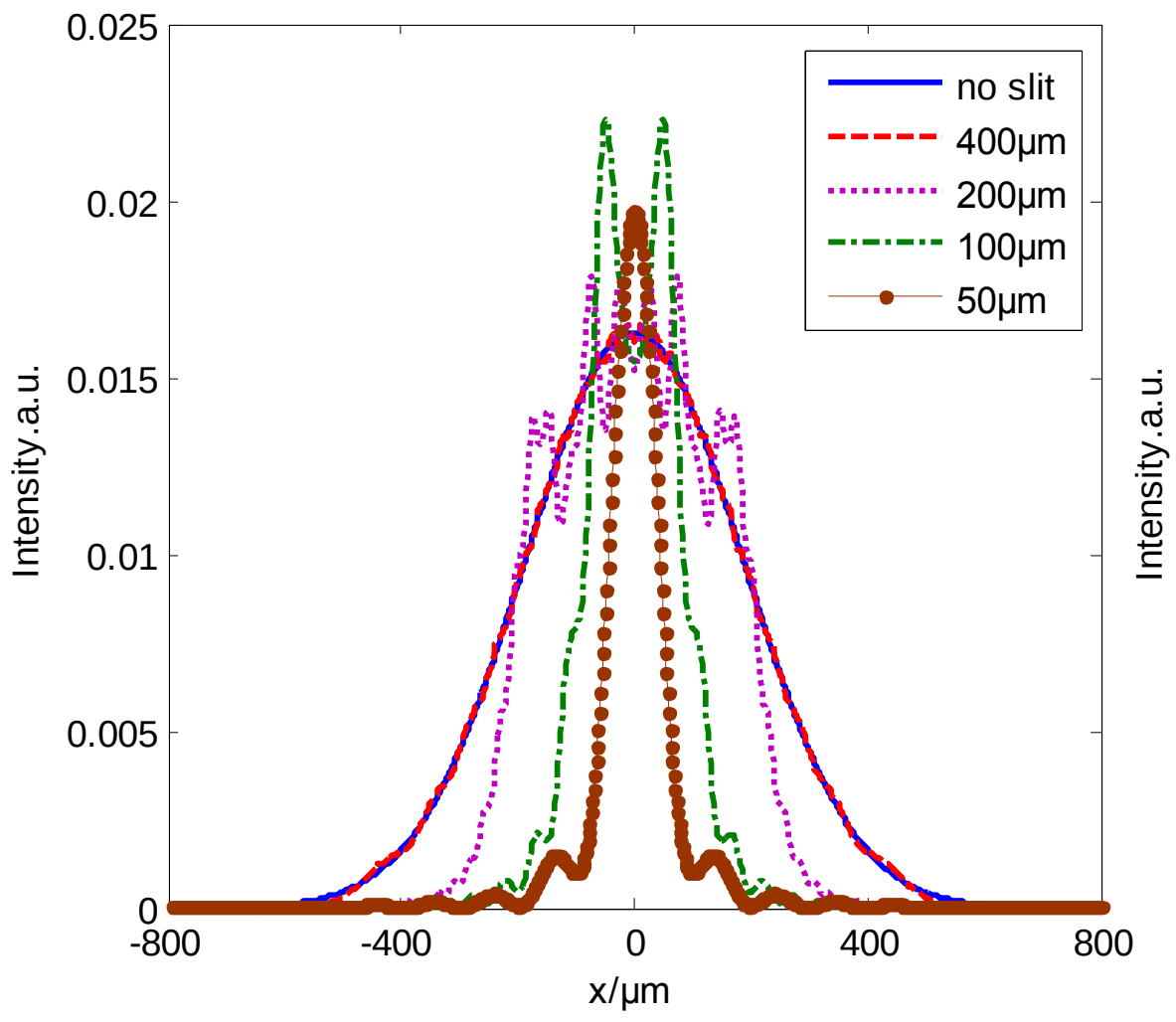

(c)

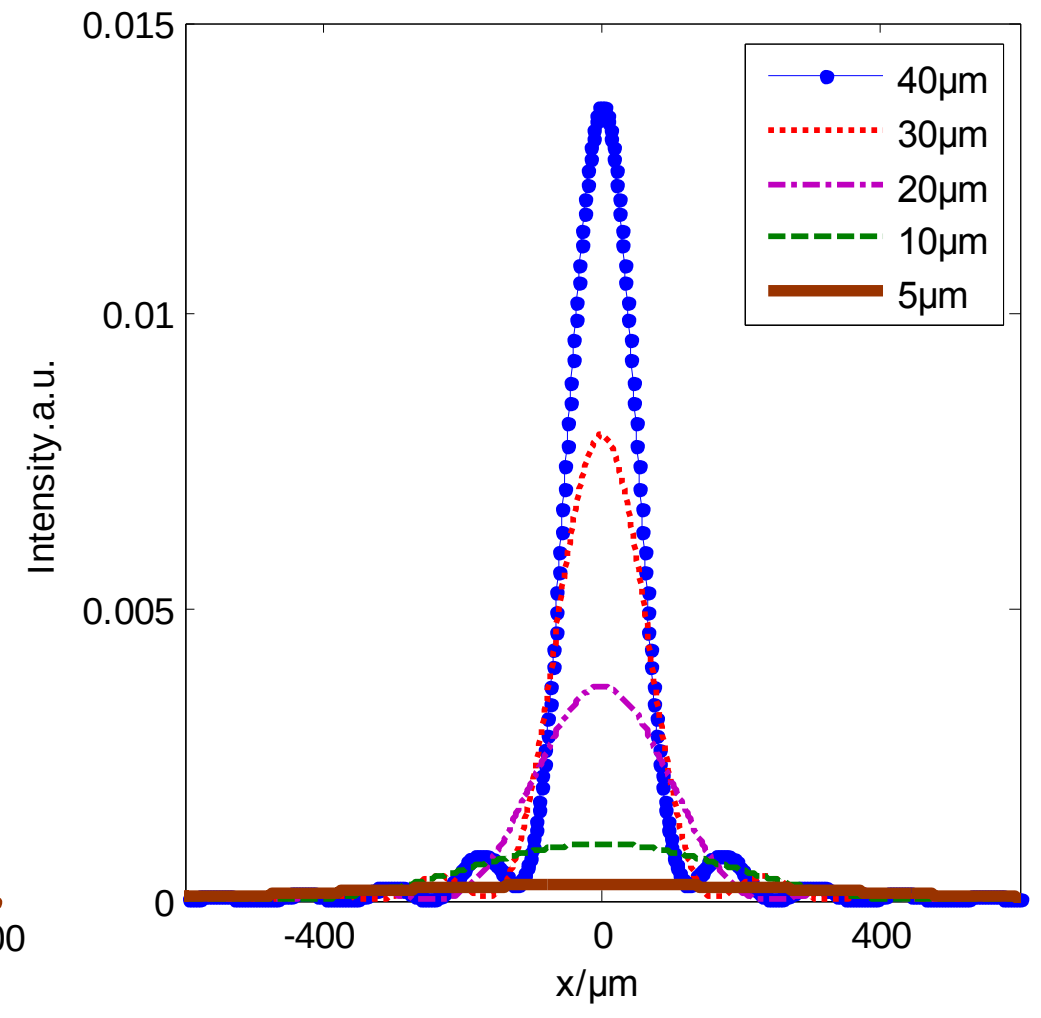

(d)

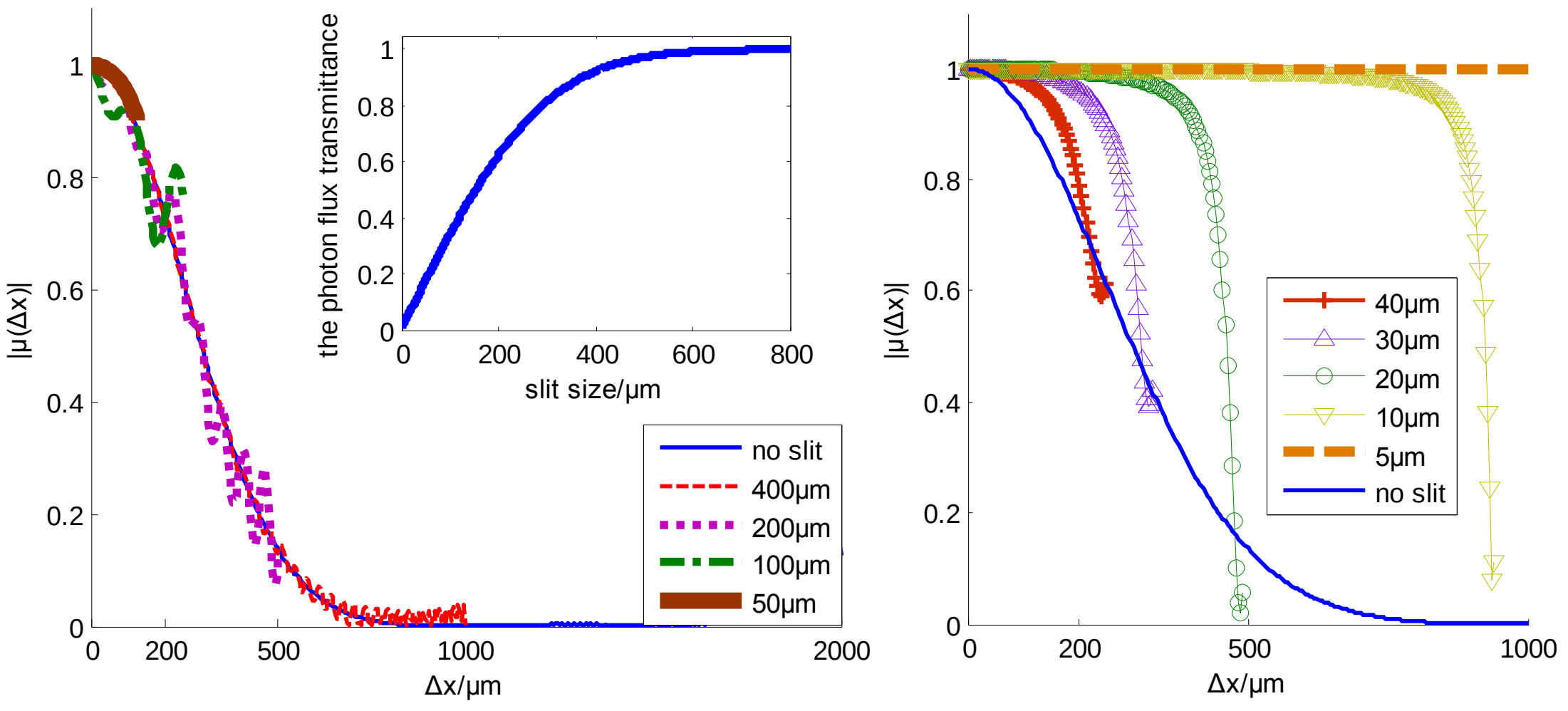

(e)

Fig. 2. The single slit diffraction intensity distribution of (a, b) the eight lowest modes and (c, d) only the fundamental mode in the observation plane. (e) The modulus of the spectral degree of coherence $|\mu(\Delta x)|$ as a function of the separation $\Delta x$ in the observation plane. Simulations are performed with no slit and different slit size of 400 μm, 200 μm, 100 μm, 50 μm, 40 μm, 30 μm, 20 μm, 10 μm, 5 μm. Inset: The photon flux transmittance as a function of the slit size.

### 3.3 Partially coherent beam focusing with an ideal thin lens

Focusing optics such as refractive lenses, waveguides, total reflection mirrors, multilayer mirrors, multilayer Laue lenses and Fresnel zone plates have been used to focus hard X-ray beams [38]. And some computer simulation methods have been developed for modeling and evaluating the focusing performance of the focusing optics [23, 26, 32, 33, 38, 39]. Historically the approach for simulating and analyzing the focusing of the synchrotron beams has evolved from neglecting the spatial coherence (geometrical optics) or treating the synchrotron source as a totally coherent point source [32, 38] instead of considering that the wavefront is partial coherent.

This article first uses the new approach based on CMD of GSM to model the focusing of the partially coherent X-ray beam with an ideal thin lens. The lens is positioned 30 m downstream the source and the intensity isophote maps of the eight lowest modes and only the fundamental mode are shown in Fig. 3(a, b). The beam's intensity distributions in the focal plane are demonstrated in Fig. 3(c). The modulus of the complex degree of coherence $|\mu(\Delta x)|$ of the focused beam as a function of the separation $\Delta x$ for different values of z is shown in Fig. 3(d). And we can clearly see that when the beam is focusing and defocusing, the coherence length of the beam is correspondingly decreasing and increasing, which is conformable to the Liouville's theorem [27, 28].

In order to facilitate the comparison and analysis, Fig. 3(e, f) show the normalized intensity pattern for the beam from a coherent point source focusing with the same lens and the beam's diffraction-limited intensity distribution in the focal plane, while this traditional method has been widely used to evaluate the focusing performance of mirrors. From the comparison of Fig. 3(a, b, c, d) and Fig. 3(e, f), we can clearly see that the new approach based on CMD of GSM can exactly model the focusing performance of the X-ray beam from the synchrotron source rather than a point source, and it can also demonstrate the coherence properties of the focusing beam. Thus, this novel method is proved to be more reasonable when compared with the traditional method.

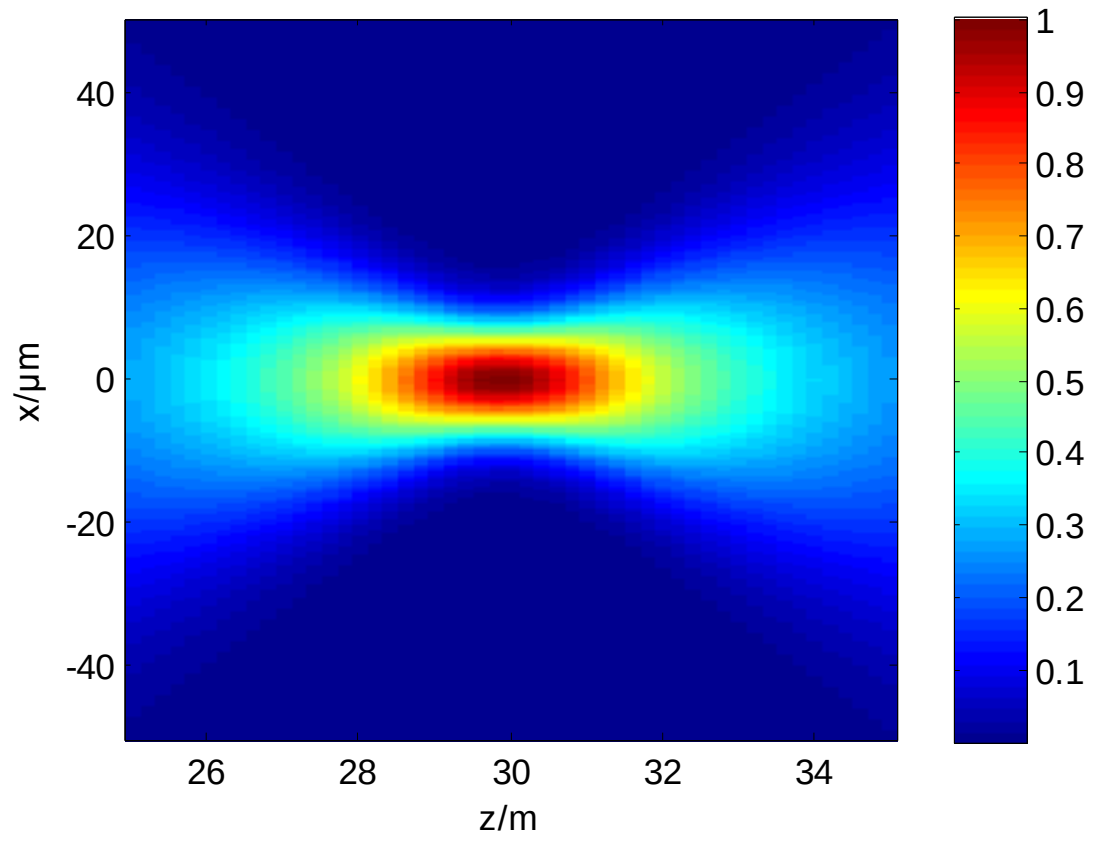

(a)

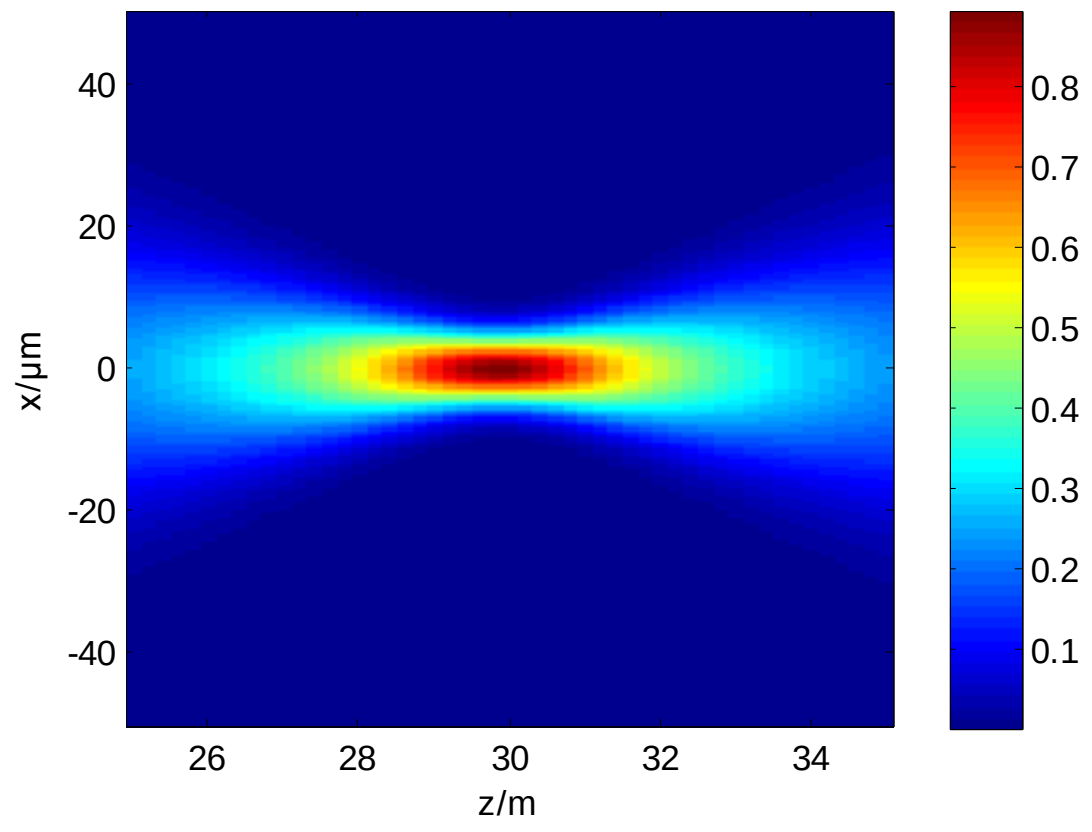

(b)

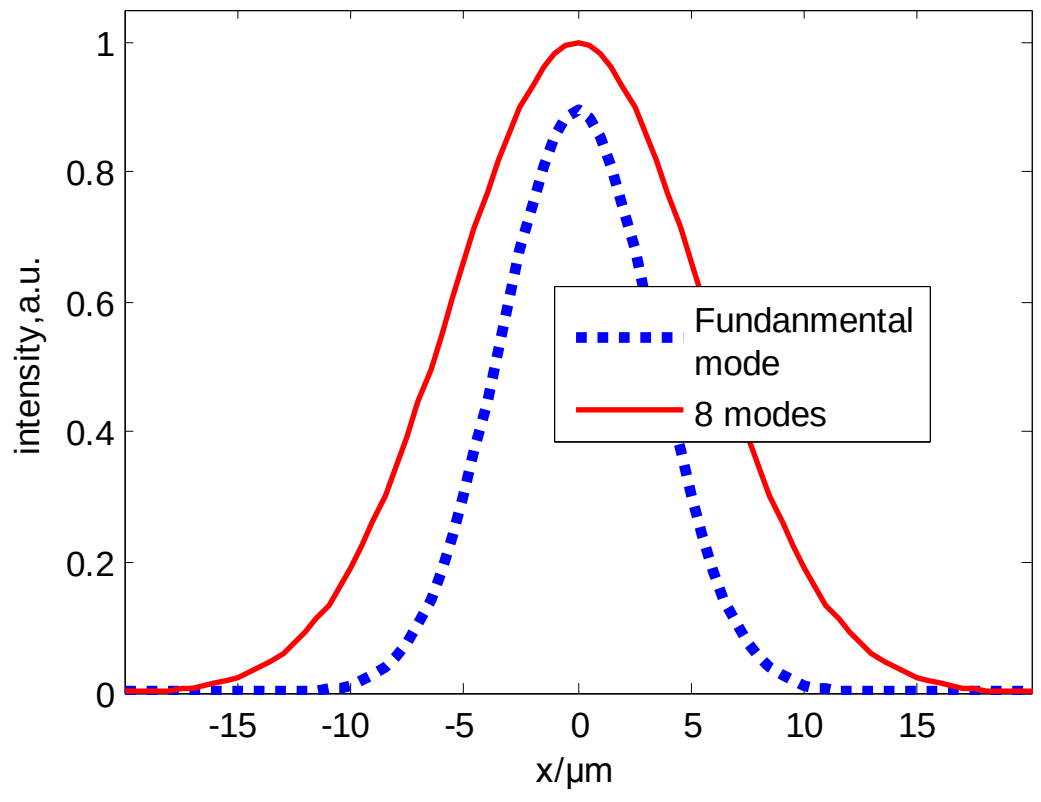

(c)

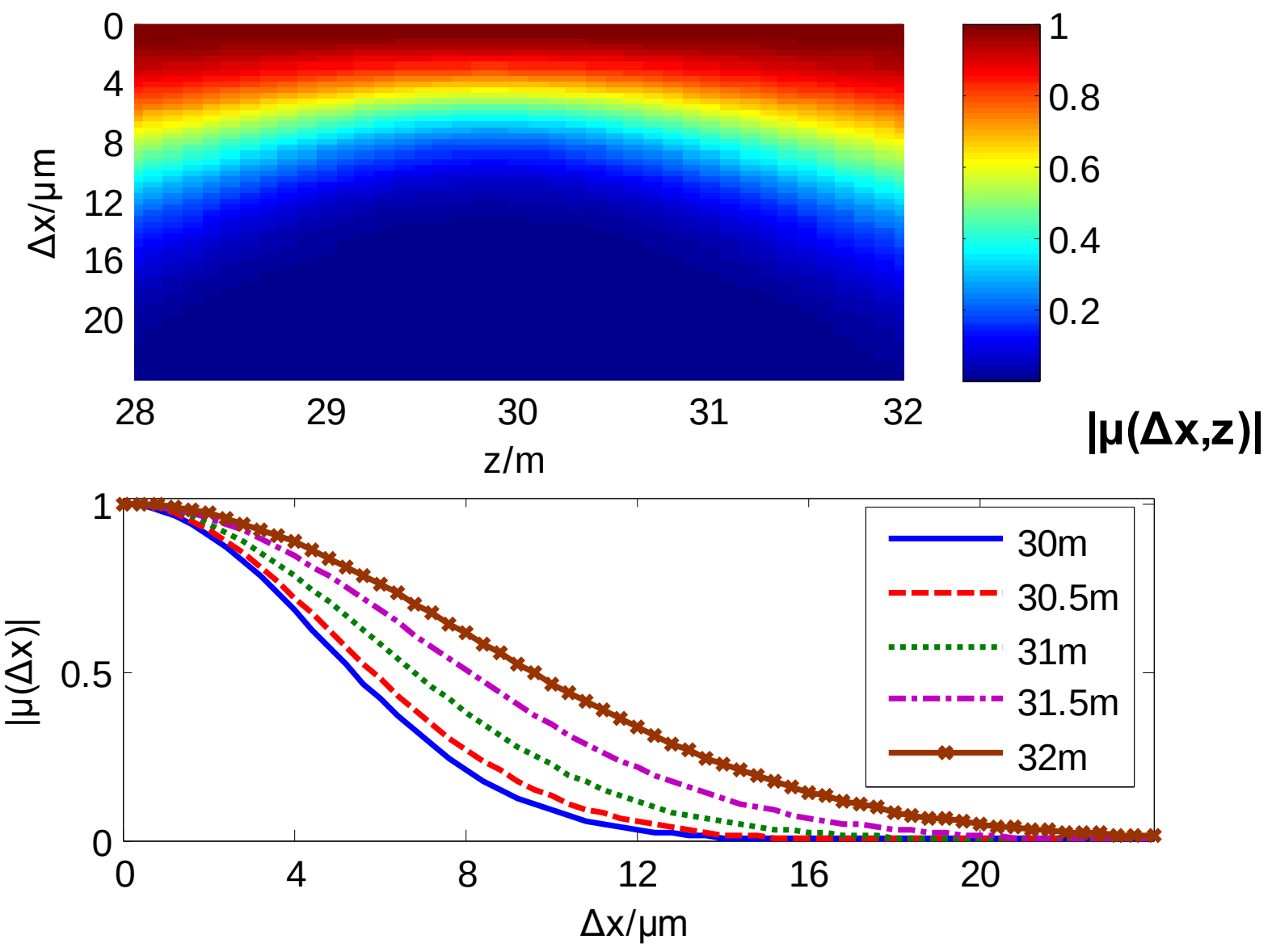
Δx/μm
0
4
8
12
16
20
28
29
30
31
32
z/m
1
0.8
0.6
0.4
0.2
|μ(Δx,z)|
|μ(Δx)|
1
0.5
0
0
4
8
12
16
20
Δx/μm
30m
30.5m
31m
31.5m
32m

(d)

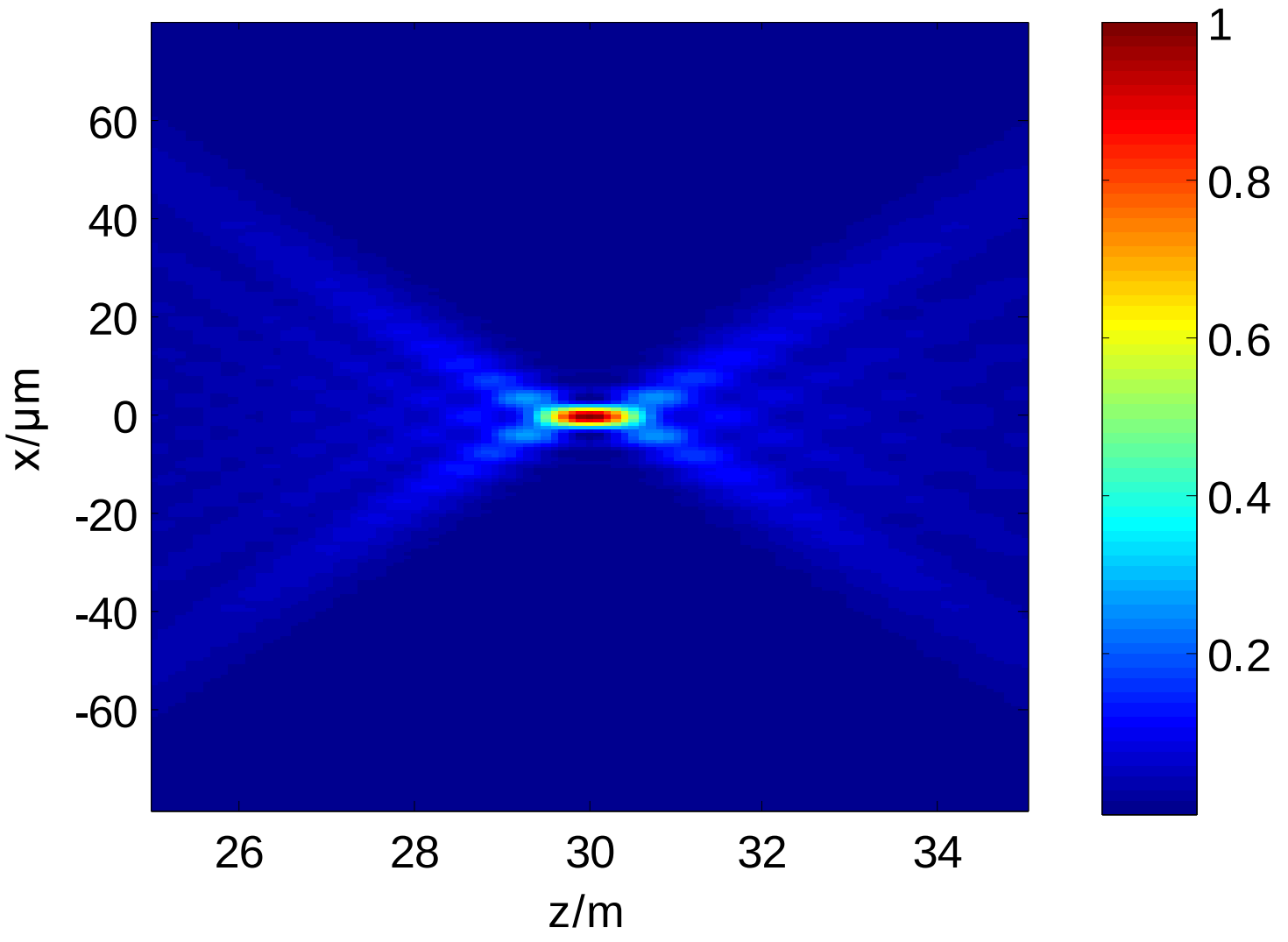

(e)

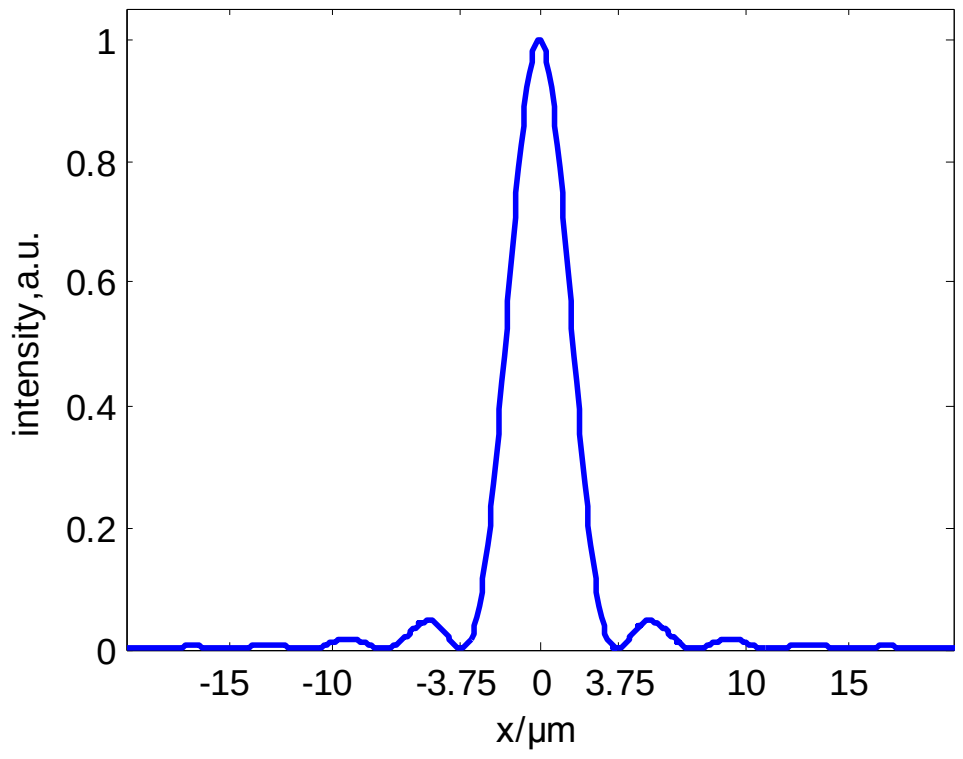

(f)

Fig. 3. The intensity isophote map of (a) the eight lowest modes and (b) only the fundamental mode for the beam focusing with an ideal thin lens. (c) The beam's intensity distribution in the focal plane. (d) The modulus of the complex degree of coherence $|\mu(\Delta x)|$ as a function of separation $\Delta x$ for different distances z (e) Isophote map showing the normalized intensity pattern for the beam from a point source focusing with an ideal thin lens. (f) The beam's diffraction-limited intensity distribution in the focal plane.

Over the past 15 years, a great class of coherence-based experiments has emerged as new tools and careful planning of these experimental is also required, which means that a better understanding of partially coherent hard X-ray beam propagation is the key for exploiting these new methods' full performance. Although some simulation methods have been developed for calculating the beam properties, but most of them can not demonstrate the coherence properties of the beam at the sample position. As the coherence properties of the hard X-ray beam at the sample position is an essential prerequisite for coherent experiments, one needs to know the effect of focusing optics on the coherence of the beam and whether the coherence can be usefully transported to the experimental apparatus. So, this new approach can be used to characterize the lateral coherence of the focusing beam at the sample position which is of vital importance in the coherent experiments. And this method is also expected to be further used for advanced predicting the coherent X-ray wavefront focal properties when taking into account the astigmation, height deviations and vibrations of focusing mirrors.

3.4 Young double slit modeling

Knowledge of the X-ray beam spatial coherence is required for the appropriate planning of the coherence-based experiments and data reduction. Quantitative characterization of the spatial coherence at two points is determined by the quality of fringe patterns generated by the interference of radiation from these two points [7]. Quantitative measurements of spatial coherence are also important in the design of experiments. Some groups have used the classical Young's experiment to measure the fringe visibility as a function of slit separations [12-15].

In this section, the Young double slit experiment is modeled and the geometry is shown in Fig. 4(a). The slit is positioned 20 m downstream the lens and the incident beam is collimated by setting the focal length of lens to 30 m as shown in Fig. 4(b). The slit size is 20 μm. Using the geometry described in Fig. 4(a), we calculate the intensity distributions of interference fringes. Fig. 5 shows the interference pattern versus different slit separations, and the patterns are obtained at different distance L between the slit and the detector position.

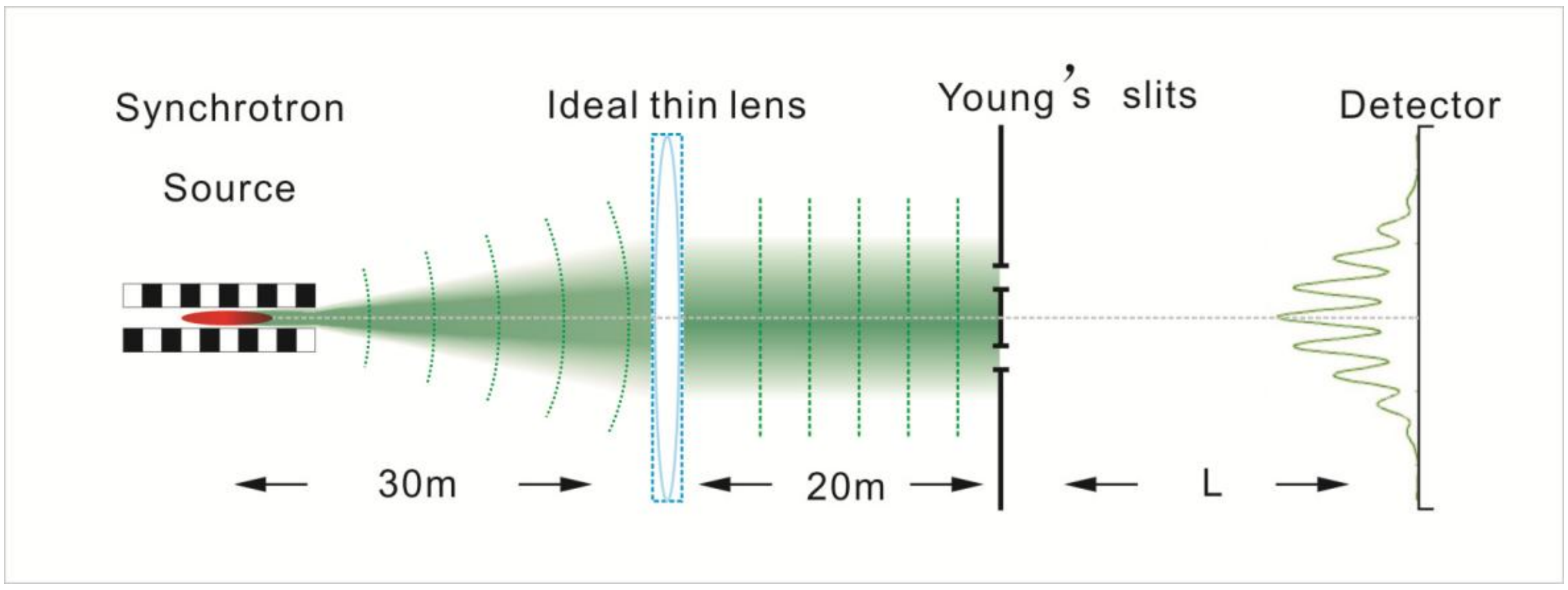

(a)

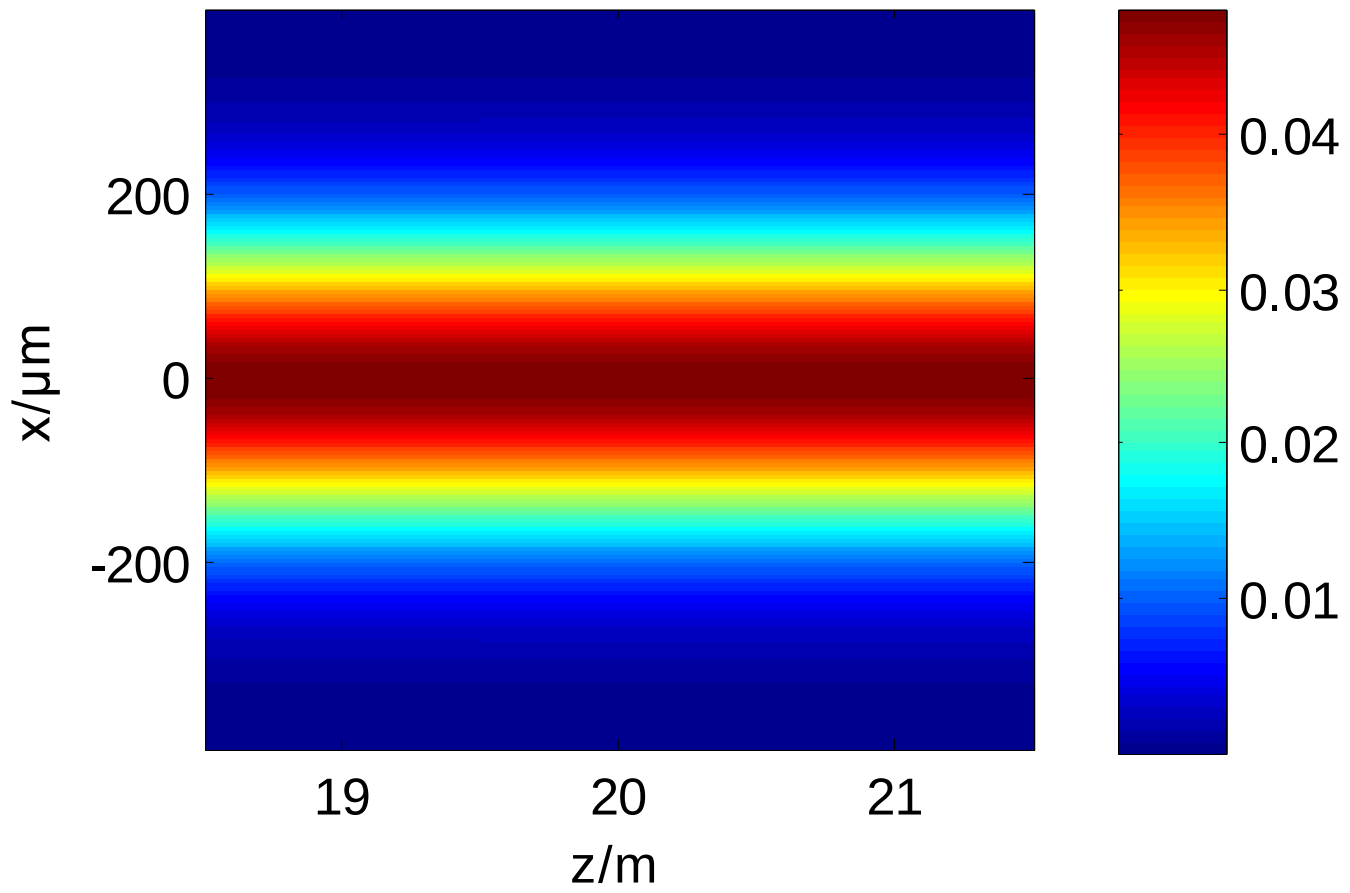

(b)

Fig. 4. (a) Double Slit diffraction geometry (slit size=20 μm). (b) Beam collimation

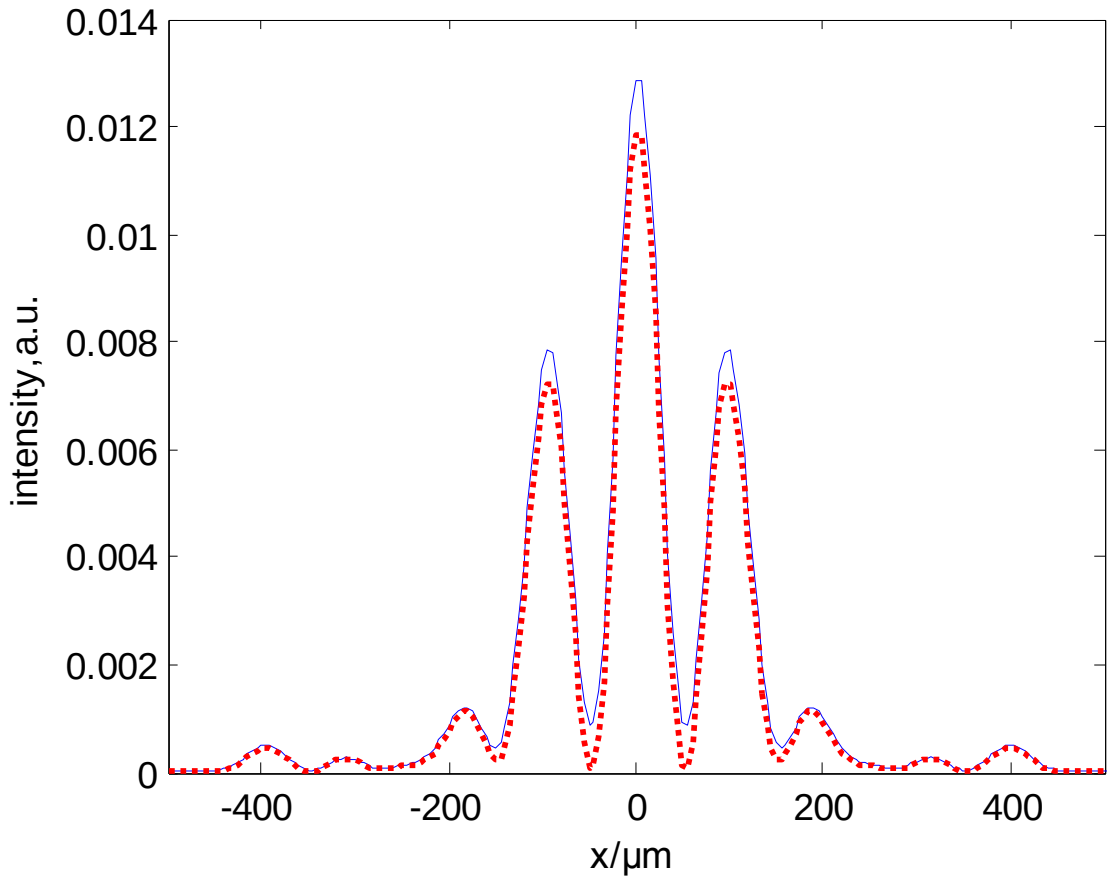

(a)

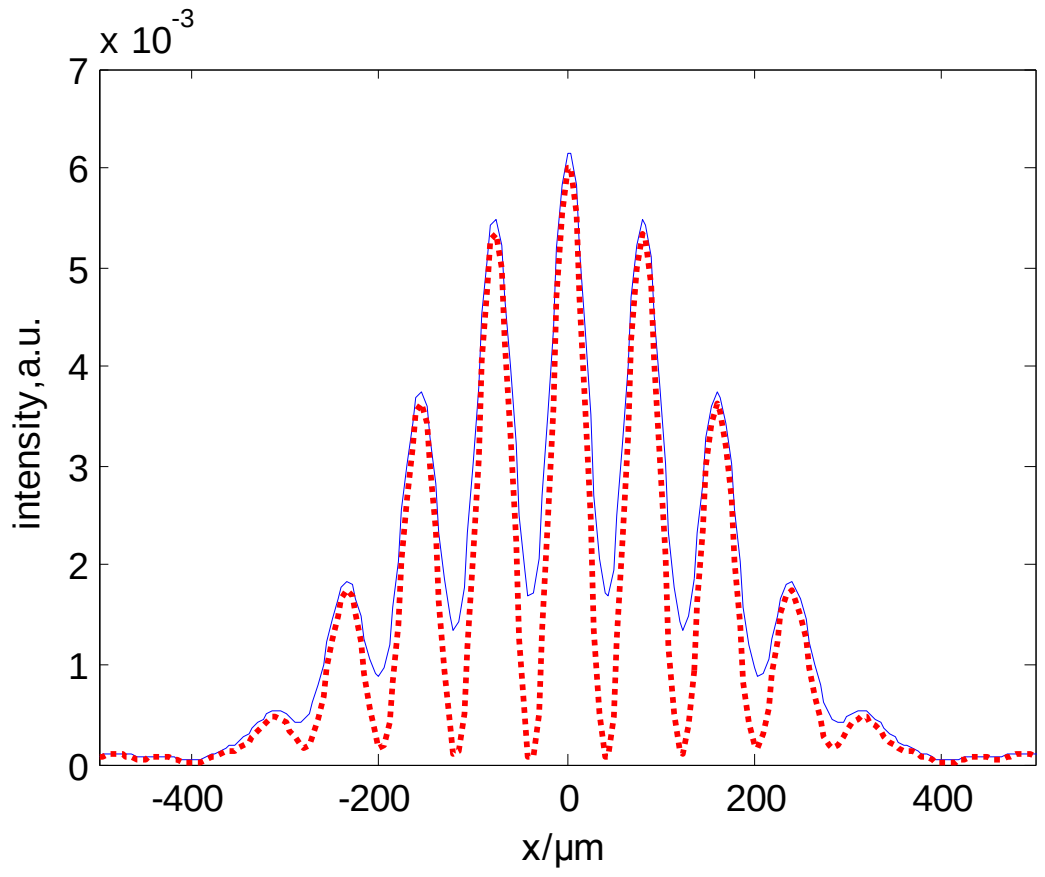

(b)

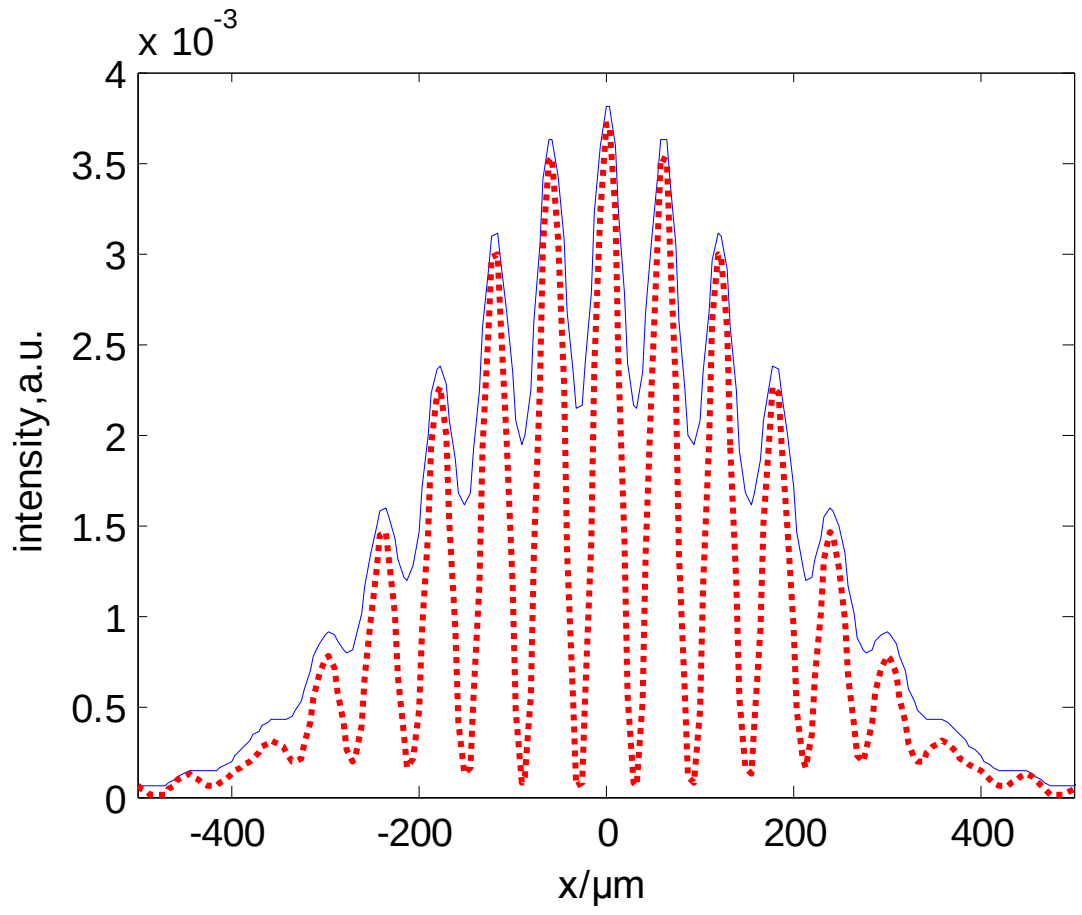

(c)

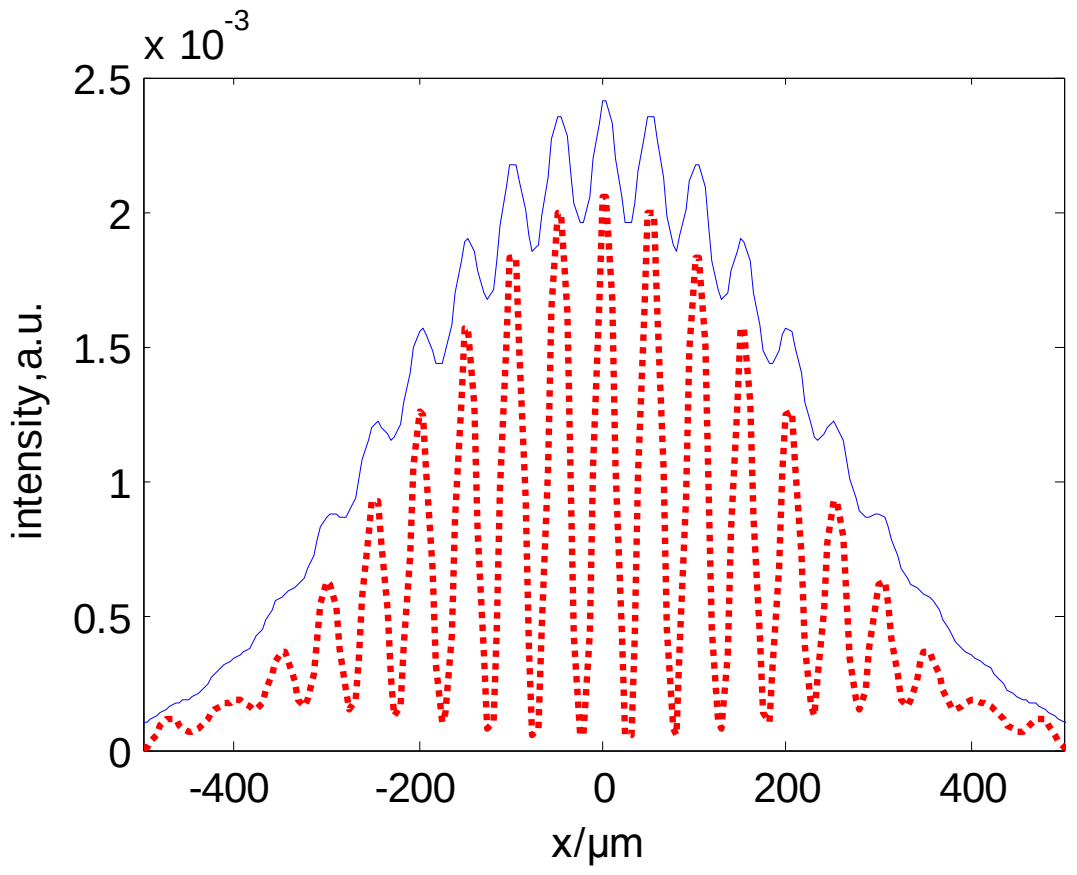

(d)

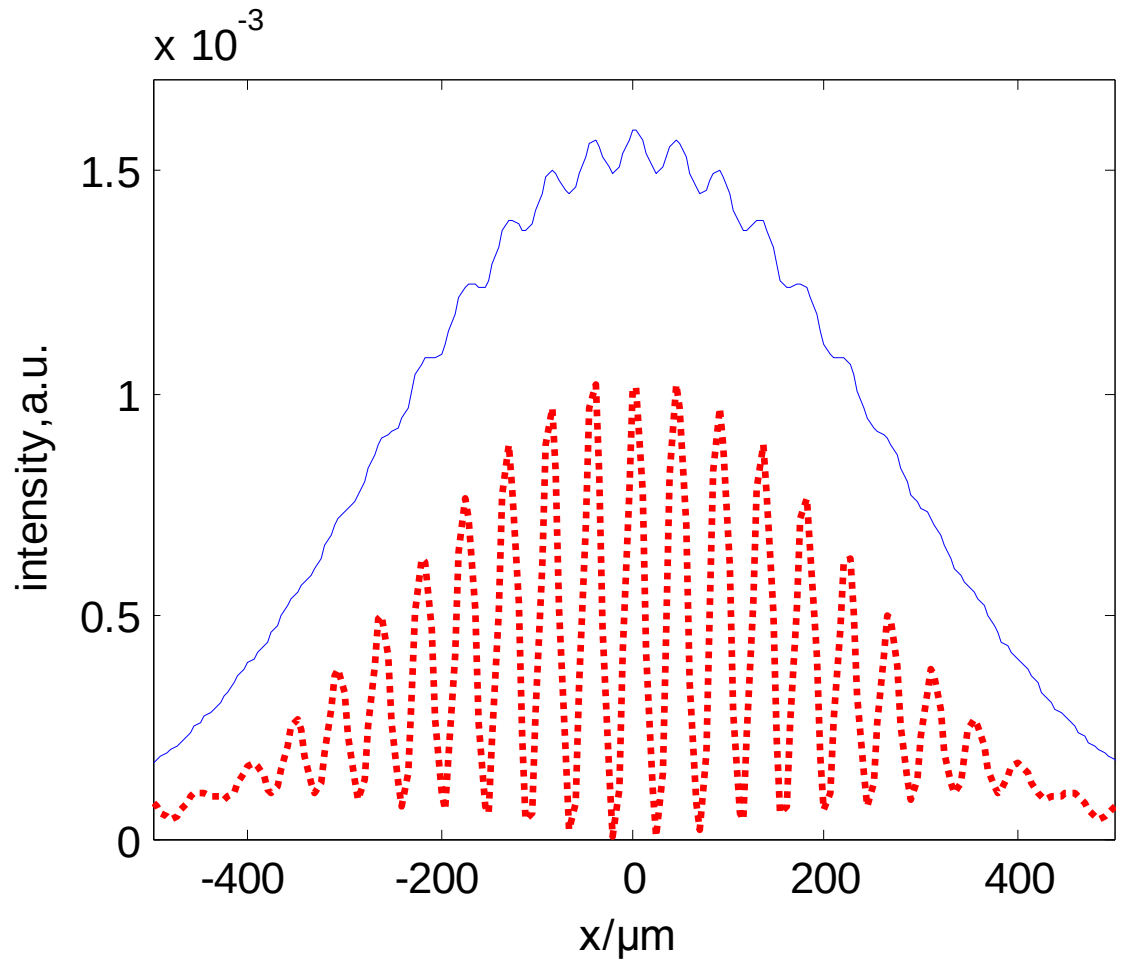

(e)

Fig. 5. Young Double slit interference pattern of the eight lowest modes (blue solid line) and only the fundamental mode (red dashed line) versus different slit separations Δx (the patterns were obtained at different distance L between the slit and the detector position). (a) Δx=50 μm, L=50 m; (b) Δx=100 μm, L=80 m; (c) Δx=150 μm, L=90 m; (d) Δx=200 μm, L=100 m; (e) Δx=250 μm, L=110 m.

The fringe visibility is defined as $V = \frac{I_{max} - I_{min}}{I_{max} + I_{min}}$, where $I_{max}$ and $I_{min}$ are the intensities at the maximum and minimum of the fringes [34]. As is shown in Table 2, the values of the fringe visibility from Fig. 5 are nearly the same as the modulus of the complex degree of coherence demonstrated in Fig. 1(c). And this also follows the function $V = \mu$, while the data's slight difference may be caused by the slit which is 20 μm rather than an ideal point.

Table 2. The modulus of the complex degree of coherence demonstrated in Fig. 1(c) and the values of the fringe visibility getting from the Fig. 5 versus different slit separations

| Slit separation | 50 μm | 100 μm | 150 μm | 200 μm | 250 μm |
|---|---|---|---|---|---|
| Results in Fig. 1(c) | 0.8684 | 0.5680 | 0.2795 | 0.1041 | 0.0291 |
| Young Double Slit modeling | 0.8719 | 0.5715 | 0.2801 | 0.1051 | 0.0308 |

## 4. Conclusions

The Gaussian Schell model's coherent mode decomposition can effectively describe the partially coherent synchrotron radiation. This approach has been applied to characterize the synchrotron source and calculate the correlation properties of the hard X-ray beams at different distances from the source.

In this paper, we extend this new powerful approach to simulate the coherence properties and intensity distributions of the hard X-ray beams passing through different optical elements. Since slit spatial filtering has usually been used to achieve spatial coherence from undulator radiation, we simulate the propagation of the X-ray beams transmitted through an ideal slit and analyze how the coherence properties and intensity distributions of the beam are changed. When the slit size gradually decreases, the transmitted beam size decreases first and then increases after arriving at a minimum value. Correspondingly, the values of $|\mu(\Delta x)|$ in the transmitted central beam which will be illuminated on the sample are enhanced significantly for smaller slits, and the correlations of the transmitted beam gradually get higher. In addition, we can decrease the slit size to get the beam with higher degree of coherence, but at the cost of photon flux. This analysis can help the beam users to count the cost as they change the slit size for more coherent beams. And it can also help the beamline designer to optimize the slits' position and size for coherent-based experiments.

This novel method can also be used for predicting the X-ray focusing performance of focusing optics in the beamline. In this paper, we investigate the properties of partially coherent X-ray beam focused by an ideal thin lens. We present the simulated intensity profiles, isophotes and the coherence properties around the focal plane for partially coherent synchrotron source. And the similar intensity patterns are also simulated for coherent illumination by a point source which has been used to evaluate the focusing performance of mirrors. This novel method is more reasonable when compared with the traditional method for both the intensity distributions and coherence properties can be obtained. And as the coherence properties of the hard X-ray beam at the sample position is an essential prerequisite for coherent experiments, this new approach can be used to characterize the lateral coherence of the focusing beam. This method is also expected to be further used in analyzing some influencing factors on the focusing performance of hard X-ray beams, such as astigmation, height deviations and vibrations of focusing mirrors. And we also model the Young double slit experiment for validating the academic analysis. Since GSM is a more reasonable model for hard X-ray synchrotron radiation, the modeling of young's double slit experiment provides the basis for the future hard X-ray beam's coherence measurements.

In our future work, we would like to extend this new approach to calculate the coherence properties of X-ray beams passing through imperfect optical elements and analyze the "decoherence" phenomenon.